\documentclass[twocolumn]{aastex63}
\pdfoutput=1
\usepackage{xcolor}
\usepackage{makecell}
\usepackage{pbox}

\shorttitle{Turbulence on galactic nuclear rings}
\shortauthors{Salas et al.}

\newcommand{\x}{\color{blue}}

\newcommand{\msun}{M$_{\odot}$}
\newcommand{\gadget}{Gadget2}
\newcommand{\e}[1]{$\times10^{#1}$}

\newcommand{\nt}{$N_t$}
\newcommand{\eturb}{$\Delta E_{in}$}
\newcommand{\splash}{\texttt{SPLASH}}

\begin{document}

\title{The effects of turbulence on galactic nuclear rings}

\correspondingauthor{Jesus~M.~Salas}
\email{jesusms@ucla.edu}

\author{Jesus~M.~Salas}
\affiliation{Dept. of Physics \& Astronomy, University of California, Los Angeles, CA, 90095, USA}

\author{Smadar~Naoz}
\affiliation{Dept. of Physics \& Astronomy, University of California, Los Angeles, CA, 90095, USA}
\affiliation{Mani L. Bhaumik Institute for Theoretical Physics, Department of Physics and Astronomy, University of California, Los Angeles, CA 90095, USA}

\author{Mark~R.~Morris}
\affiliation{Dept. of Physics \& Astronomy, University of California, Los Angeles, CA, 90095, USA}

\begin{abstract}
The gas dynamics in the inner few kiloparsecs of barred galaxies often results in configurations that give rise to nuclear gas rings. 
However, the generic dynamical description of the formation of  galactic  nuclear  rings  does not  take  into account the  effects of thermal pressure and turbulence.
Here we perform 3D hydrodynamic simulations of gas in a galactic barred potential out to a radius of $2$~kpc. We include self-gravity and a large-scale turbulence-driving module. We study how the formation of gaseous nuclear rings is affected by changing the bar pattern speed and the strength of the turbulence, and compare the results to simulations with a relatively high sound speed. We use two values for the bar pattern speed that have been discussed in the literature for our Milky Way Galaxy (40 and 63 km s$^{-1}$ kpc$^{-1}$). Our results show that turbulence produces broader and smaller nuclear rings, and enhances gas migration towards the inner few pc of the galaxy, compared to simulations without turbulence. 
\end{abstract}
\section{Introduction} \label{sec:intro}
Nuclear gaseous rings are a common morphological feature of barred galaxies \citep[e.g.,][]{Buta1996FCPh...17...95B,Knapen2005A&A...429..141K,Comeron2010MNRAS.402.2462C,Comeron2013A&A...555L...4C,Buta2017MNRAS.471.4027B,Buta2017MNRAS.470.3819B}. Stellar bars introduce non-axisymmetric torques which produce morphological substructures in the gaseous medium, such as a pair of dust lanes at the leading side of the bar, and a nuclear gaseous ring near the center \citep[e.g.,][]{Sanders1976ApJ...209...53S,Roberts1979ApJ...233...67R,Athanassoula1992MNRAS.259..345A,Buta1996FCPh...17...95B,Martini2003ApJS..146..353M,Martini2003bApJ...589..774M}. These nuclear rings can serve as a gas reservoir for the accretion disk that surrounds the supermassive black hole (SMBH) that is present at the centre of most galaxies.

Similarly, our Galaxy has such a ring of radius $\sim100-150$~pc, corresponding to the densest and most massive part of the Central Molecular Zone (CMZ, \citealt{Morris1996,Molinari2011,Kruijssen2015MNRAS.447.1059K,Henshaw2016MNRAS.457.2675H}). The CMZ is thought to be created and fed from the outside by the Galactic bar. According  to the most widely accepted theory of galactic dynamics, the gas initially settles into $X_1$ orbits, which occur between the corotation radius and the inner Lindblad resonance (ILR) of the bar potential \citep[e.g.,][]{Binney1991MNRAS.252..210B}. As inwardly migrating gas approaches the ILR, there is an innermost stable $X_1$ orbit inside of which the orbits become self-intersecting. The gas compresses and shocks near the edges of these orbits, loses angular momentum and descends onto $X_2$ orbits, which are closed and elongated orbits that have their long axes oriented perpendicular to the bar (\citealt{Binney1991MNRAS.252..210B,Athanassoula1992MNRAS.259..345A,Jenkins1994MNRAS.270..703J,Gerhard1996IAUS..169...79G}). The shocks along the innermost $X_1$ orbit are presumed responsible for compressing the gas into molecular form, and the accumulated molecular gas on $X_2$ orbits comprises the observed CMZ \citep[e.g.,][]{Binney1991MNRAS.252..210B}. However, it is unclear how fast molecular gas is transported further in toward the central few parsecs, and which mechanisms are responsible for its transport.

The generic dynamical description of the formation of a gaseous nuclear ring does not take into account the effects of thermal pressure. For example, \cite{Patsis2000A&A...358...45P,Kim2012ApJ...747...60K,Sormani2015aMNRAS.449.2421S} and \cite{Sormani2018aMNRAS.481....2S}, showed that, for a given underlying gravitational potential, the size and morphology of nuclear rings depend on the sound speed of the gaseous medium. Furthermore, it has been shown that the size and location of nuclear rings are also loosely related to the location of the ILR, and thus to the bar pattern speed, although the predicted location is more accurate for strongly barred potentials (e.g., \citealt{Buta1996FCPh...17...95B,Sormani2015bMNRAS.451.3437S,Sormani2018aMNRAS.481....2S}).

Relatively high gas temperatures ($70$-$100$~K) are one of the key properties of CMZ clouds, and there is evidence showing that the gas is  kept  warm  by  the  dissipation of turbulence \citep{Immer2016,Ginsburg2016}. Furthermore, the large turbulent velocity dispersion within the CMZ must be responsible for supporting the gas against gravitational collapse, since the thermal pressure of the gas  would be insufficient. This motivates the need to balance the effects of self-gravity. Generally, the effects of turbulence on galactic-scale simulations have been investigated by using momentum and energy injection from supernova (SN) explosions, which are known to drive turbulence in the interstellar medium (ISM, e.g., \citealt{Norman1996ApJ...467..280N,MacLow2004RvMP...76..125M,Joung2006ApJ...653.1266J}). However, we take a different approach from previous studies by driving the turbulence via a Fourier forcing module, based on the methods by \cite{Stone1998ApJ...508L..99S} and \cite{MacLow1999}. The details of this method are described in \cite{Salas2019}.

In this work we perform simulations of gas residing in the central few kiloparsecs of a barred galaxy, using 3D smoothed-particle hydrodynamics (SPH). Our main goal is to apply our turbulence driving method to simulations of gaseous nuclear rings that include self-gravity (i.e., the mutual gravitational interactions between the SPH particles), and to compare how the effects of forced turbulence differ from the effects of thermal pressure in both low and high sound speed simulations. 

This paper is organized as follows: we briefly describe our numerical methods in Section \ref{sec:numerical_methods}. We describe our main results in Section \ref{sec:results}. Finally, we discuss our results and approximations and present some concluding remarks in Section \ref{sec:discussion}.
\section{Numerical Methods}\label{sec:numerical_methods}

We used the N-body/SPH code \gadget\ \citep{Springel2005}, which is based on the tree-Particle Mesh method for computing gravitational forces and on the SPH method for solving the Euler equations of hydrodynamics. The smoothing length of each particle in the gas is fully adaptive down to a set minimum of $0.001$~pc. \gadget\ employs an entropy formulation of SPH, as outlined in \cite{Springel2002}, with the smoothing lengths defined to ensure a fixed mass (i.e., fixed number of particles) within the smoothing kernel volume (set at N$_{neigh} = 64$). The code adopts the Monaghan-Balsara form of artificial viscosity \citep{Monaghan1983,Balsara1995}, which is regulated by the parameter $\alpha_{MB}$, set to $0.75$. 

We modified the standard version of \gadget\ to include turbulence driving and the gravitational potential of a Milky Way-type galaxy. We describe these modifications below. 
\subsection{The galactic potential} \label{subsec:potential}
To calculate the gravitational potential, we use the density profile from \cite{Zhao1994AJ....108.2154Z}:
\begin{equation}\label{eq:potential}
\rho(r,\theta,\phi) = \rho_0 \left ( \frac{r}{r_0}\right)^{-p} [1 + Y(\theta, \phi)] \ ,
\end{equation}
which is a modified version of the prolate bar density profile introduced by \cite{Binney1991MNRAS.252..210B}. The gravitational potential then has the form:
\begin{equation}\label{eq:potential}
\Phi(r,\theta,\phi) = 4\pi G \rho_0 r_0^2 \left(\frac{r}{r_0}\right)^\alpha P(\theta,\phi) \ ,
\end{equation}
where $(r,\theta,\phi)$ are spherical coordinates fixed on the rotating bar\footnote{The coordinate $r = \sqrt{x^2+y^2+z^2}$, where $x,y,z$ are the standard Cartesian coordinates. The supermassive black hole would be at $r = 0$, the bar's major axis is aligned with the $x$-axis, and the $z$ axis represents the vertical direction, with the galactic plane at $z=0$}, $\alpha = 2-p$, and $P(\theta,\phi)$ is the associated Legendre function, which can be written as:
\begin{equation}
P(\theta,\phi) = \frac{1}{\alpha (1+\alpha)} - \frac{Y(\theta,\phi)}{(2-\alpha)(3+\alpha)} \ ,
\end{equation}
and $Y(\theta,\phi)$ is a linear combination of spherical harmonic functions of the
$l= 2$, $m= 0,2$ modes:
\begin{equation}
Y(\theta,\phi) = -b_{20} P_{20} (\cos\theta) + b_{22}P_{22}(\cos\theta)\cos2\phi \ .
\end{equation}
The parameter $b_{20}$ determines the degree of oblateness/prolateness
while $b_{22}$ determines the degree of non-axisymmetry. Motivated by the previous work of \cite{Kim2011ApJ...735L..11K}, and more recently of \cite{Gallego2017}\footnote{We note that there is a negative sign misprint in \cite{Kim2011ApJ...735L..11K} (their Equation 2) and in \cite{Gallego2017} (their associated Legendre function).}, we use the parameters: $\alpha = 0.25$, $b_{20}=0.3$, $b_{22} =0.1$, $\rho_0 =40$ \msun\ pc$^{-3}$ and $r_0 = 100$ pc. Given these parameters, a bar with axis ratios of [1: 0.74: 0.65] is obtained for the isodensity surface that intersects points [$x$ = 0, $y$ = $\pm200$~pc, $z$ = 0].
Enclosed masses inside $200$~pc and $1000$~pc are $10^9$ \msun\ and $7\times10^9$ \msun, respectively.

In addition to the gravitational force due to the potential above, we performed the computation in a reference frame rotating with the bar, and therefore included centrifugal and Coriolis forces. We compare two different values for the bar pattern speed: a ``fast'' bar with $\Omega_{bar} = 63$ km s$^{-1}$ kpc$^{-1}$, which has been adopted by previous studies \citep[e.g.,][]{Kim2011ApJ...735L..11K,Sormani2015aMNRAS.449.2421S, Krumholz2015MNRAS.453..739K}, and a ``slow'' bar with $\Omega_{bar} = 40$ km s$^{-1}$ kpc$^{-1}$, which is the most recent estimate of the pattern speed of the Galactic bar \citep{Hawthorn2016ARAA..54..529B,Portail2017MNRAS.465.1621P}. 

\subsection{Turbulence Driving}\label{subsec:turbulence}

Supersonic turbulence occurs over a wide range of length scales in the interstellar medium, especially within molecular clouds. The importance of turbulence in modulating star formation in the interstellar medium was highlighted recently by a combination of numerical and analytical studies  \citep[e.g.,][]{Krumholz2005ApJ...630..250K,Burkhart2018ApJ...863..118B}. Furthermore, turbulence in the CMZ seems to greatly influence its thermal structure and star formation rate \citep[e.g.,][]{Kruijssen2014MNRAS.440.3370K}. 

Numerical simulations have shown that turbulence decays quickly, within a few dynamical timescales \citep[e.g.,][]{Stone1998ApJ...508L..99S,MacLow1999}. Since observations indicate high turbulent velocity dispersions in the CMZ clouds \citep{Morris1996}, turbulence then must be driven by some physical stirring mechanism, e.g., magnetic fields, secular gas instabilities, feedback ejecta, etc. However, the main driving mechanism for turbulence in the CMZ has not yet been definitively identified (see \cite{Kruijssen2014MNRAS.440.3370K} for a discussion of possible sources of turbulence).

Simulations of turbulence-driven gas are often 
employed in studies of the interstellar medium and star formation \citep[e.g.,][]{Stone1998ApJ...508L..99S,MacLow1998PhRvL..80.2754M,Krumholz2005ApJ...630..250K,Burkhart2009ApJ...693..250B,Federrath2010AA...512A..81F}. Typically, this is achieved by a Fourier forcing module, which can be modelled with a spatially static pattern in which the amplitude is adjusted in time \citep{Stone1998ApJ...508L..99S,MacLow1999}. Other studies employ a forcing module that can vary both in time and space \citep[e.g.,][]{Padoan2004PhRvL..92s1102P,Schmidt2006,Federrath2010AA...512A..81F}.

In the case of galaxy simulations, driven turbulence is mimicked by injecting energy due to SN. For example, \citet{Kim2011ApJ...735L..11K,Emsellem2015MNRAS.446.2468E,Shin2017ApJ...841...74S,Seo2019ApJ...872....5S,Armillotta2019MNRAS.490.4401A}, and \cite{Tress2020MNRAS.499.4455T} have modelled turbulence by using star formation and SN feedback models. In general, these models depend on underlying assumptions regarding star formation rates, SN energies and injection rates. Furthermore, recent studies have demonstrated that the different choices of SN feedback model (including the underlying physical processes driving the feedback) produce significant differences in morphology, density, etc, of the simulated galaxies \citep[e.g.,][]{Scannapieco2012MNRAS.423.1726S,Rosdahl2017MNRAS.466...11R,Keller2020arXiv200403608K}.

In order to avoid relying on a particular physical mechanism, we adopt a Fourier forcing module, which has the advantage of being independent of the source of turbulence. Our turbulence treatment is based on the method described by \cite{MacLow1999}, in which a turbulent velocity field is drawn from a spatially static pattern having a power spectrum $P(k)\propto k^{-n}$, where $k$ is the wavenumber. 
We describe this turbulence model, as well as the performance tests conducted to show its effectiveness, in more detail in \cite{Salas2019}. For completeness we summarize the key factors of the algorithm here. We create a library of 10 spatially static turbulent velocity fields (in the form of cubic lattices, or grids). Each lattice is created using fast Fourier transforms inside a $128^3$ box, resulting in a realization of a turbulent velocity field with power spectrum $P(k) \propto k^{-4}$, from $k=2$ to $k=128$. We fill the volume of our simulation domain ($4$~kpc per side) with $64^3$ cubic lattices, each drawn randomly from our library. Each lattice is given a physical size of $64$~pc per side. Thus, turbulence is driven at scales of $64/2$ = $32$~pc (for $k=2$) to $64/128$ = $0.5$~pc (for $k=128$). We use tri-linear interpolation to calculate the velocity ``kicks'' given to every gas particle inside each lattice. The amplitude of the velocity kicks is adjusted in time to maintain a constant energy input. Finally, all of the turbulent lattices are changed randomly every time the driving is performed. 

\begin{table*}

\begin{tabular}{|c|c|c|c|c|c|}
\hline
 Test name & Turbulence & $c_s$ & Self gravity & Pattern speed \\
           &  & (km s$^{-1}$) &   & (km s$^{-1}$ kpc$^{-1}$) \\           
\hline
SBLSP   &  None   & 0.6 & No  &    40                \\ \hline
SBHSP   &  None   & 10  & Yes &    40                \\ \hline
SBLT    &  Low & 0.6 & Yes &    40               \\  
&  (\eturb = $10^{47}$~ergs, \nt = 100)  &  & &  \\
\hline

SBHT    &  High & 0.6 & Yes &    40               \\
&  (\eturb = $10^{47}$~ergs, \nt = 2)  &  & &  \\
\hline
\hline
FBLSP   &  None   & 0.6 & No  &    63        \\ \hline
FBHSP   &  None   & 10  & Yes &    63        \\ \hline
FBLT    &  Low  & 0.6 & Yes &    63           \\ 
&  (\eturb = $10^{47}$~ergs, \nt = 100)  &  & & \\ \hline
FBHT    &  High & 0.6 & Yes &    63           \\
&  (\eturb = $10^{47}$~ergs, \nt = 2)  &  & &  \\\hline
\end{tabular} 
\caption{Summary of all tests. SB and FB stand for ``slow bar'' and ``fast bar'', respectively. LSP and HSP stand for ``low sound speed'' and ``high sound speed'', respectively. HT and LT stand for ``high turbulence'' and ``low turbulence'', respectively. The low turbulence models correspond to injecting \eturb = $10^{47}$ ergs of energy every \nt = 100 timesteps, and the high turbulence models correspond to injecting \eturb = $10^{47}$ ergs of energy every \nt = 2 timesteps (see Section \ref{subsec:turbulence}).}
\label{table:models}

\end{table*}

Similarly, this turbulence driving algorithm was recently implemented in SPH simulations of the Circumnuclear Disk \citep[CND,][]{Dinh2021}. By adjusting the injection rate and the sizes of the turbulence grids, \cite{Dinh2021} mimicked the effects of turbulence sources with scales similar to the size of the disk. Their results demonstrate that turbulence can give rise to a long lived structure, which suggest that the CND itself may also be long lived, as opposed to be a transient structure, which has been suggested by previous studies. 

Our turbulence implementation contains two free parameters: \eturb, the total energy input per injection, and \nt, the number of timesteps between velocity ``kicks'' (the timestep is fixed in all simulations to be equal to 1000 yrs). In \cite{Salas2019}, we demonstrate that our turbulence module produces consistent results in the range $E_{turb}$ = $10^{46}-10^{50}$ ergs, which corresponds to $\sim0.01-100\%$ of the thermal energy of the system. We also show that \nt\ must be relatively low (\nt = 2-5) in order to counteract the self-gravity of the high-density gas, due to its fast free-fall time ($t_{ff} \sim 1/\sqrt{G \rho})$. In the present work, however, we expect the densities of our large-scale simulations to be much lower than those we studied in \cite{Salas2019} (and thus a larger $t_{ff}$), which allows us to consider larger values for \nt.

Here, we consider two extremes, namely a ``low turbulence'' model and a ``high turbulence'' model. This is achieved by tuning the two free parameters, $E_{turb}$ and $N_t$. For simplicity, we fix $E_{turb}$ to be $10^{47}$~ergs, and use two values for $N_t$, 100 and 2, which represent the low and high turbulence models, respectively. We expect the turbulence parameters in real galactic centers to fall somewhere between these two extremes.

\subsection{Initial conditions}\label{subsec:IC}

As a proof-of-concept, we create a simplistic model of a galactic disk consisting of an outer radius of $2$~kpc, an inner radius of 30 pc, and a Gaussian scale height of $50$~pc. The disk contains a total mass of $10^8$ \msun, with each SPH particle having a mass of $130$~\msun. The particles are initially in circular orbits, with their velocities calculated using the potential described in Section \ref{subsec:potential}. All simulations were run using an isothermal equation of state.

\section{Results} \label{sec:results}
\begin{figure*}
\includegraphics[width=1\textwidth]{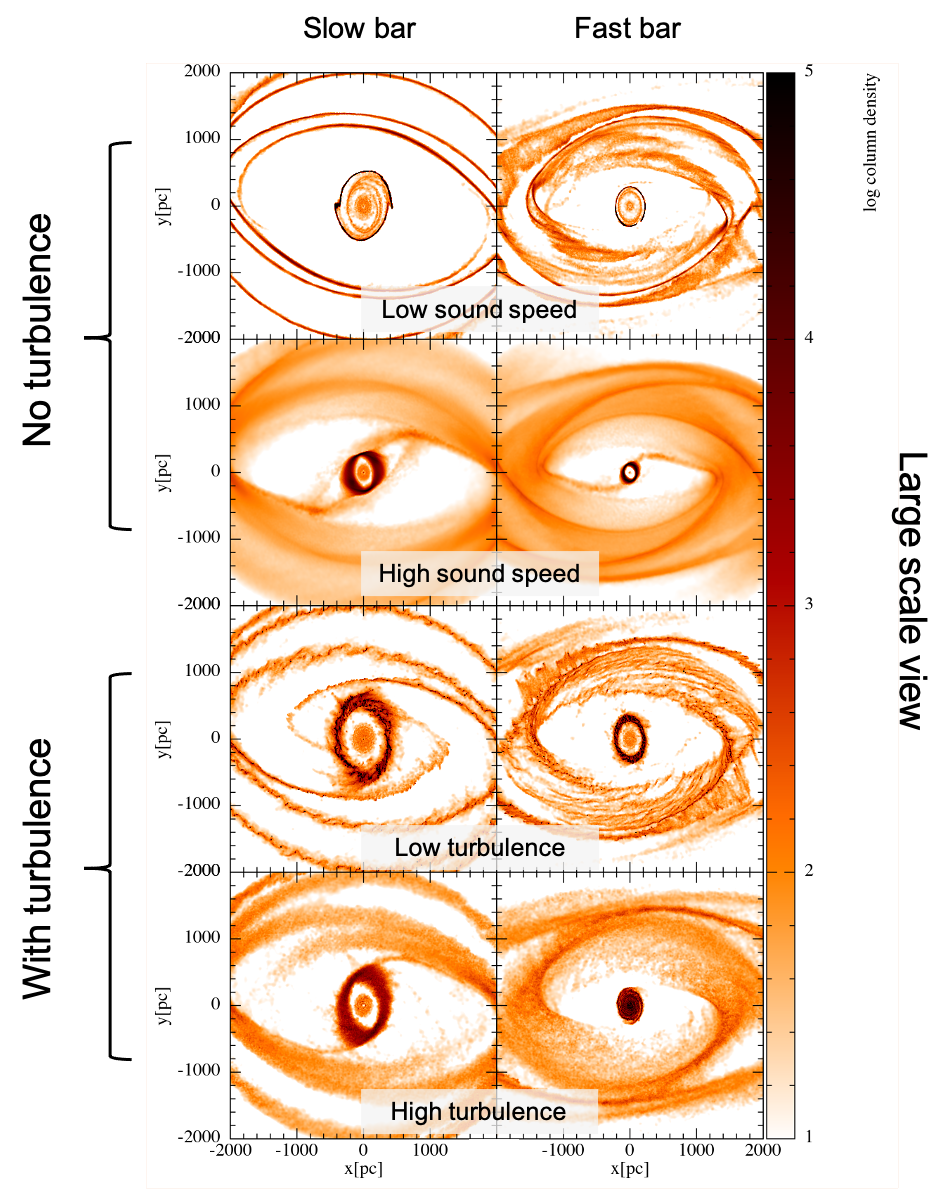}
\caption{Face-on view of the simulations at t = 200 Myrs. The slow-bar models produce a bigger ring than the fast-bar models, as expected. Furthermore, the high turbulence tests make the nuclear rings more spread out (dispersed). The long axis of the bar lies along the x-axis. } 
\label{fig:big_face_on}
\end{figure*}

\begin{figure*}
\includegraphics[width=1\textwidth]{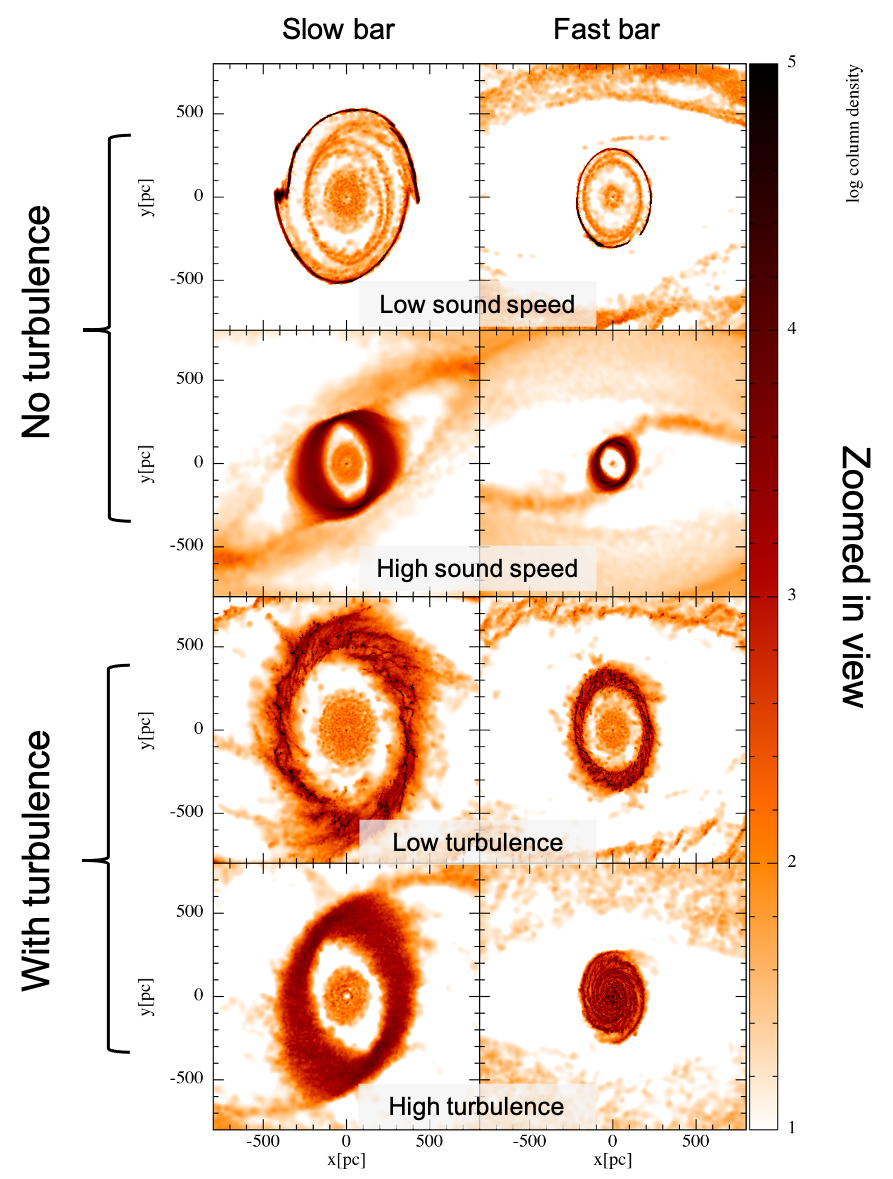}
\caption{Face-on view of the simulations at t = 200 Myrs, zoomed-in to the inner 800x800 pc. The long axis of the bar lies along the x-axis.} 
\label{fig:zoom_face_on}
\end{figure*}

\begin{figure*}
\centering
\includegraphics[width=0.8\textwidth]{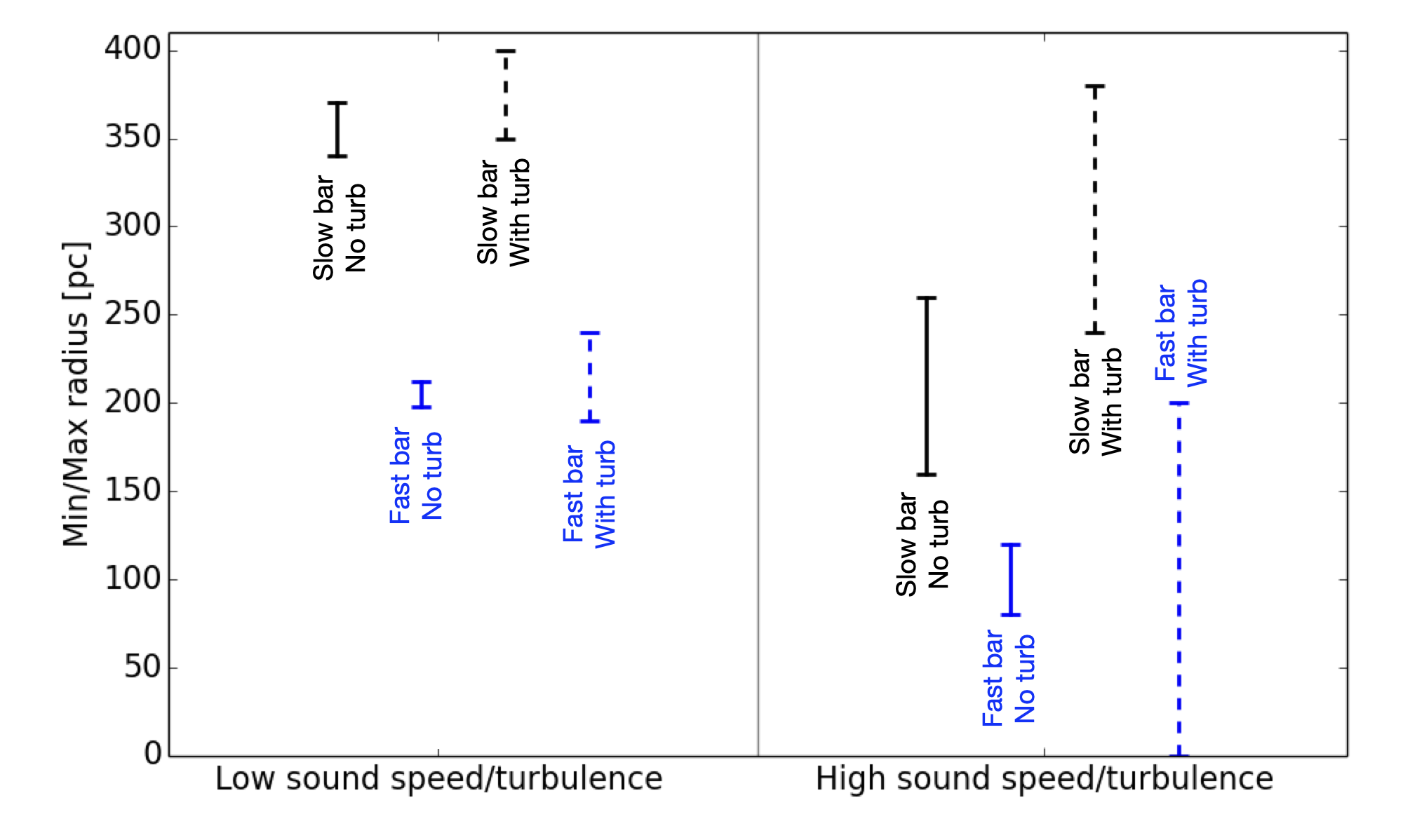}
\caption{Radial thickness comparison for the nuclear rings in all 8 simulations. For each vertical radial distance line, the points (horizontal dashes) represent the minimum and maximum radial distances of the ring (along the $+x$ axis).
Black lines indicate the slow bar tests, while blue lines indicate the fast bar tests. Solid lines indicate the tests without turbulence, and dashed lines indicate the tests with turbulence. } 
\label{fig:thick}
\end{figure*}

We performed 4 tests with turbulence driving, using the slow and fast bar pattern speed values described in Section \ref{subsec:potential}, and the low and high turbulence parameters described in Section \ref{subsec:turbulence}. The system reaches steady state by $\sim150$~Myrs, thus we ran simulations ran for $200$~Myrs, which is long enough to capture the relevant dynamics, in addition to save computational time. For comparison, we also performed 4 tests with no turbulence: two of these simulations were done using a sound speed of $c_s = 0.6$~km~s$^{-1}$ (same as with the runs with turbulence. This value corresponds to a temperature of about $100$~K, assuming the gas is primarily molecular), and the other two were done using a sound speed of $c_s = 10$~km~s$^{-1}$. Each of these was also done with a slow and fast bar. Table \ref{table:models} summarizes the parameters used in each test. The nomenclature is as follows: FB and SB correspond to ``Fast Bar" and ``Slow Bar'', respectively. LSP and HSP correspond to ``Low Sound Speed'' and ``High Sound Speed''. Similarly, LT and HT indicate ``Low Turbulence'' and ``High Turbulence'', respectively.

\subsection{Ring comparison}
Here we investigate the morphology of the rings produced by the different bar pattern speeds and the effects of turbulence.
Figure \ref{fig:big_face_on} and \ref{fig:zoom_face_on} show the final state of the gas at $t$ = 200 Myrs for all simulations performed. 
In all runs, the gas accumulates on $X_2$ orbits, forming an elongated ring. Moreover, the smallest nuclear rings were produced by the fast bar tests, as expected, since the bar pattern speed influences the location of the nuclear ring \citep[e.g.,][]{Sormani2015cMNRAS.454.1818S}. The size of a nuclear ring is related to the radius of the ILR \citep{Buta1996FCPh...17...95B,Sormani2018aMNRAS.481....2S}: increasing the pattern speed pushes the ILR inward, thus yielding a smaller ring.  

For clarity, we encapsulate the size difference between the nuclear rings in Figure \ref{fig:thick}, where we show the spread (radial thickness, i.e., the minimum and maximum radius along the $+x$ axis) of the rings in each simulation. Black lines indicate the slow bar tests, while blue lines indicate the fast bar tests. Solid lines indicate the tests without turbulence, and dashed lines indicate the tests with turbulence.

Among the simulations without turbulence, the tests with high sound speed produce smaller nuclear rings than those with low sound speed. This effect has been explored by \cite{Kim2012ApJ...747...60K} and later by \cite{Sormani2015aMNRAS.449.2421S,Sormani2018aMNRAS.481....2S}, who demonstrated that nuclear rings shrink in size with increasing sound speed. In particular, \cite{Sormani2018aMNRAS.481....2S} found that because of the high sound speed in their simulations, the thermal pressure forces are significant and lead to the development of shocks. This shocked gas is slowed down, and starts falling closer to the center than in low sound speed simulations (for more details on the effects of thermal pressure on nuclear rings, see Section 5.2 of \citealt{Sormani2018aMNRAS.481....2S}).

Meanwhile, our turbulence treatment induces a more dispersed (spread out) structure to the nuclear rings compared to the runs without turbulence (especially those with low sound speed). Similarly, the high turbulence runs produce more dispersed nuclear rings than the low turbulence (and the low sound speed) runs, which was expected. Additionally, turbulence induces a more filamentary structure in gas density, compared to the more diffuse gas density in the simulations without turbulence (as was also described in \citealt{Salas2019}).
In Figure \ref{fig:zoom_face_on} we show a zoom-in of the inner $800\times800$ pc of the simulations, where these differences can be seen more clearly. 

In particular, the simulation with a fast bar and high turbulence (FBHT) produces a nuclear ring that is completely filled and smaller than in all other turbulence runs. This effect is similar to that found in the simulations by \cite{Tress2020MNRAS.499.4455T}, which demonstrated that turbulence due to SN feedback greatly influences inflow from the CMZ to the central few parsecs. Our turbulence methodology (with a high turbulence energy injection, as in the FBHT test) captures these effects, which help drive gas from the nuclear ring to the inner few pc.

As mentioned in \cite{Salas2019}, the spreading of the nuclear rings can be attributed to the ``turbulent'' viscosity induced by our driving method. Because of this viscosity, angular momentum is transferred from the inner parts of the nuclear ring to the outer parts. The inner parts of the ring lose some angular momentum and move inward, while the outer parts gain some angular momentum and move outward. The net effect is that the rings spreads \citep[e.g.,][]{Shakura1973A&A....24..337S,Lynden1974MNRAS.168..603L,Pringle1981ARA&A..19..137P}. We estimate the turbulent viscosity induced by our driving method below.

\subsection{Turbulent viscosity}
\begin{table*}
\includegraphics[width=1\textwidth]{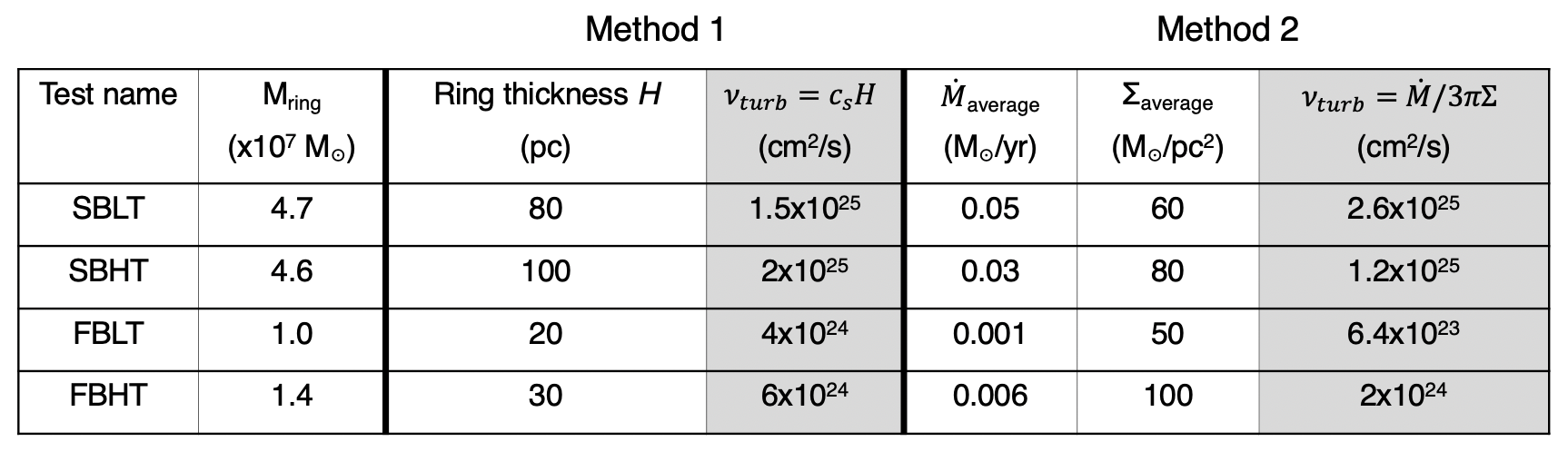}
\caption{Tabulated values used to calculate the turbulent viscosity, $\nu_{turb}$, for each of the runs with turbulence. }
\label{table:visc_table}
\end{table*}

We successfully showed in \cite{Salas2019} that we can approximate the viscosity induced by our driving module using the concept of $\alpha$-viscosity
\citep{Shakura1973A&A....24..337S,Pringle1981ARA&A..19..137P}. First, we use the standard definition:
\begin{equation} \label{eq:alpha1}
    \nu_{turb} = \alpha_\nu c_s H \ ,
\end{equation}
where $\alpha_\nu  \leq 1$ is a parameter that adjusts the strength of the viscosity, and $H$ is the vertical thickness (scale height) of the nuclear ring. We calculate the average vertical thickness $H$ using the density map at $t=200$ Myrs, and calculate $\nu_{turb}$ for each ring in the simulations with turbulence (assuming $\alpha_\nu =1$), and present them in Table \ref{table:visc_table}, shown under the label ``Method 1''. We note that the nuclear rings in the ``slow bar'' (SB) tests are vertically thicker (larger $H$) than the ``fast bar'' (FB) tests, thus leading to higher values of $\nu_{turb}$. This can be explained by the fact that a slower bar pattern speed creates bigger rings, which contain more mass (which we also indicate in Table \ref{table:visc_table}, shown as $M_{ring}$), and thus  receive a larger fraction of the injected turbulent energy than the smaller rings.

Alternatively, as we showed in  \cite{Salas2019}, $\nu_{turb}$ can also be estimated by using the mass accretion rate due to $\alpha$-viscosity \citep{Shakura1973A&A....24..337S,Pringle1981ARA&A..19..137P}:
\begin{equation} \label{eq:alpha2}
    \nu_{turb} = \frac{\dot{M}}{3 \pi \Sigma}
\end{equation} \ ,
where $\dot{M}$ is the mass accretion rate and $\Sigma$ is the surface density. 

\begin{figure*}
\centering
\includegraphics[width=1\textwidth]{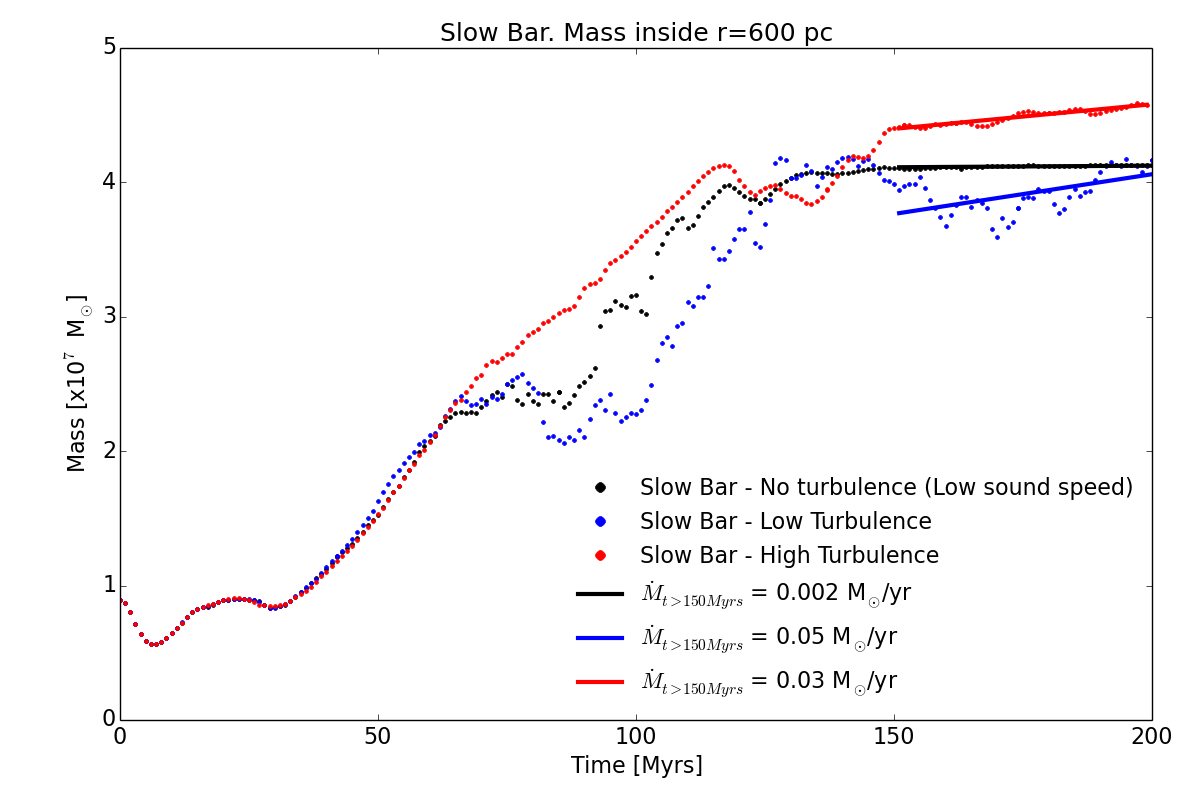}
\includegraphics[width=1\textwidth]{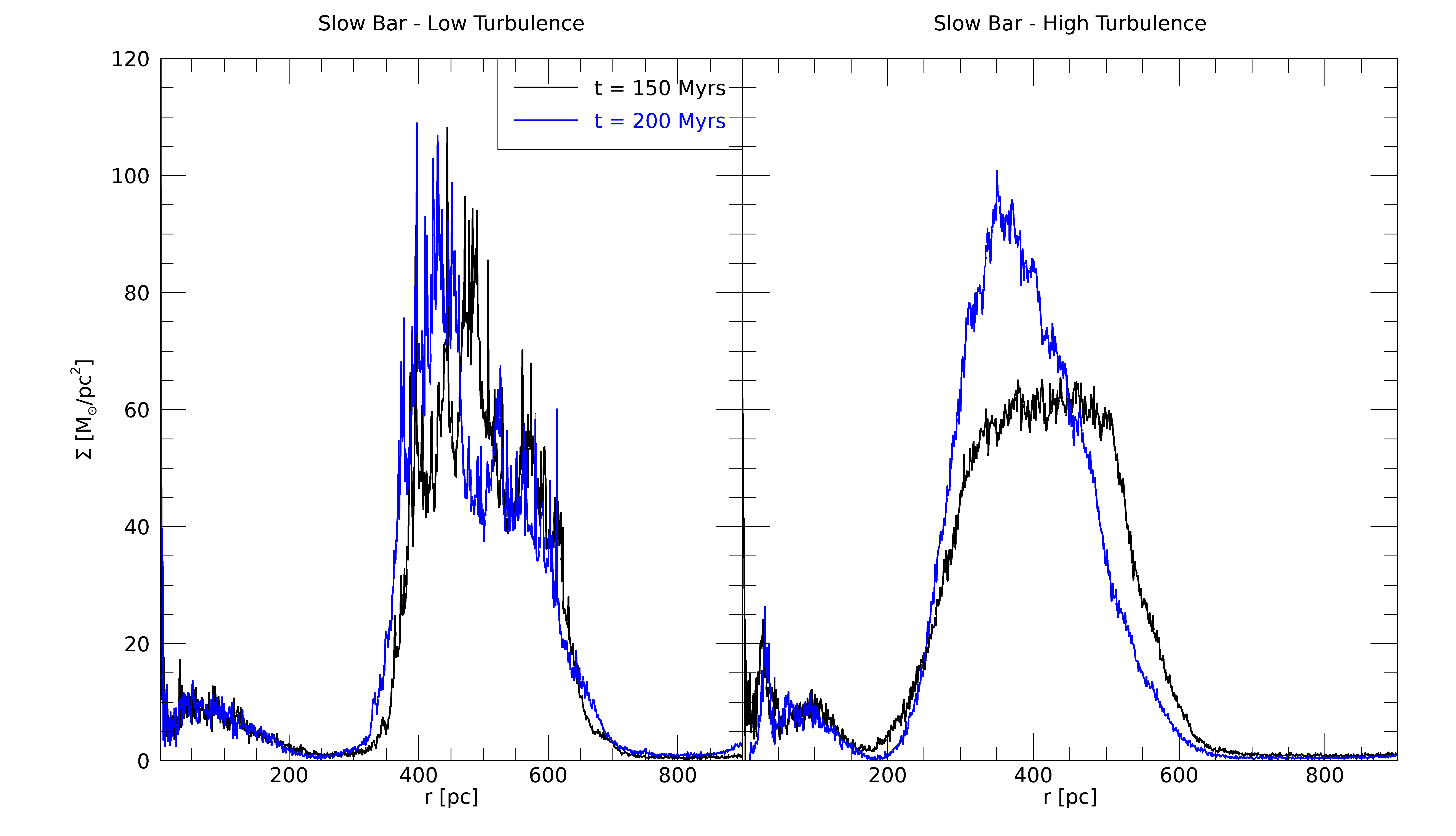}
\caption{Top: mass inside a radius of $r=600$~pc as a function of time. The black points represent the slow bar, no turbulence simulation (SBLSP). Similarly, the blue and red points represent the low turbulence (SBLT) and high turbulence (SBHT) simulations, respectively. We approximate the mass evolution after $150$~Myrs (i.e., after steady-state has been reached) as a straight line, and compute the slope, $\dot{M}$. Bottom: plots of surface density ($\Sigma$) versus radius for the SBLT (bottom left) and SBHT (bottom right) simulations. The black and blue lines correspond to $t=150$ and $200$~Myrs, respectively.   } 
\label{fig:slow_bar_plots}
\end{figure*}

\begin{figure*}
\centering
\includegraphics[width=1\textwidth]{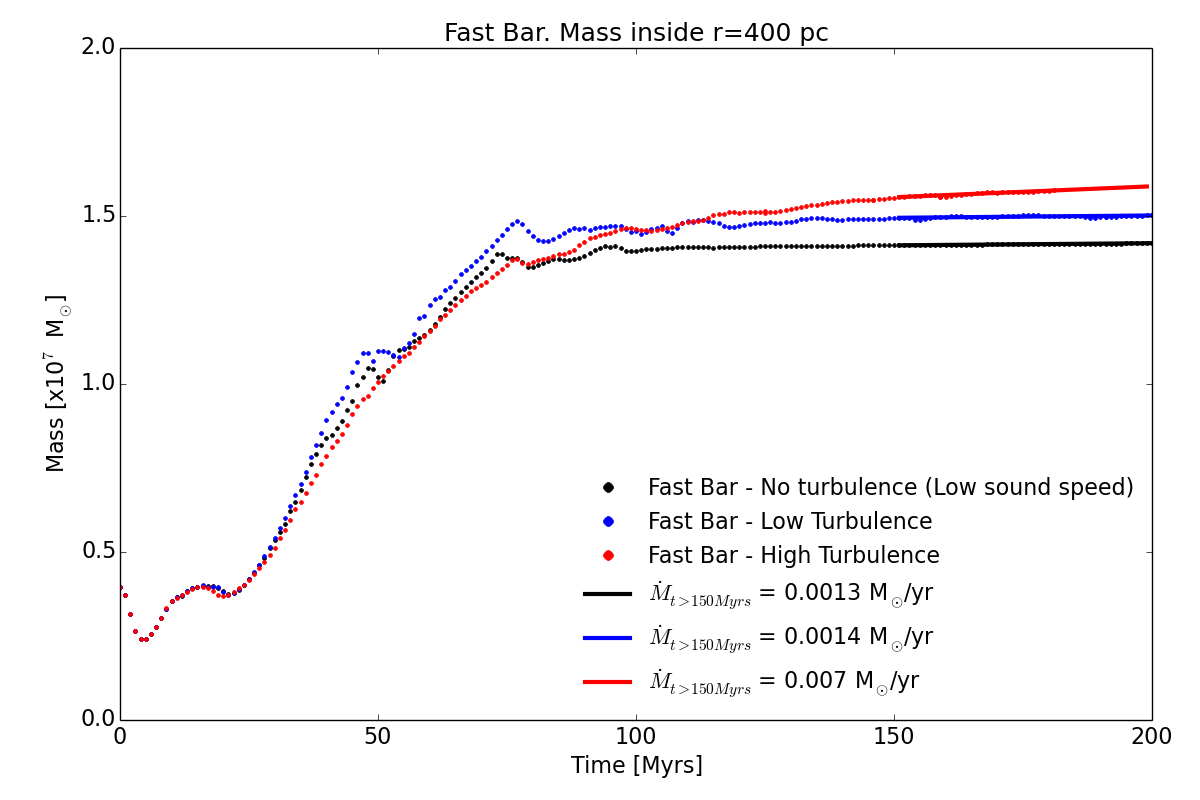}
\includegraphics[width=1\textwidth]{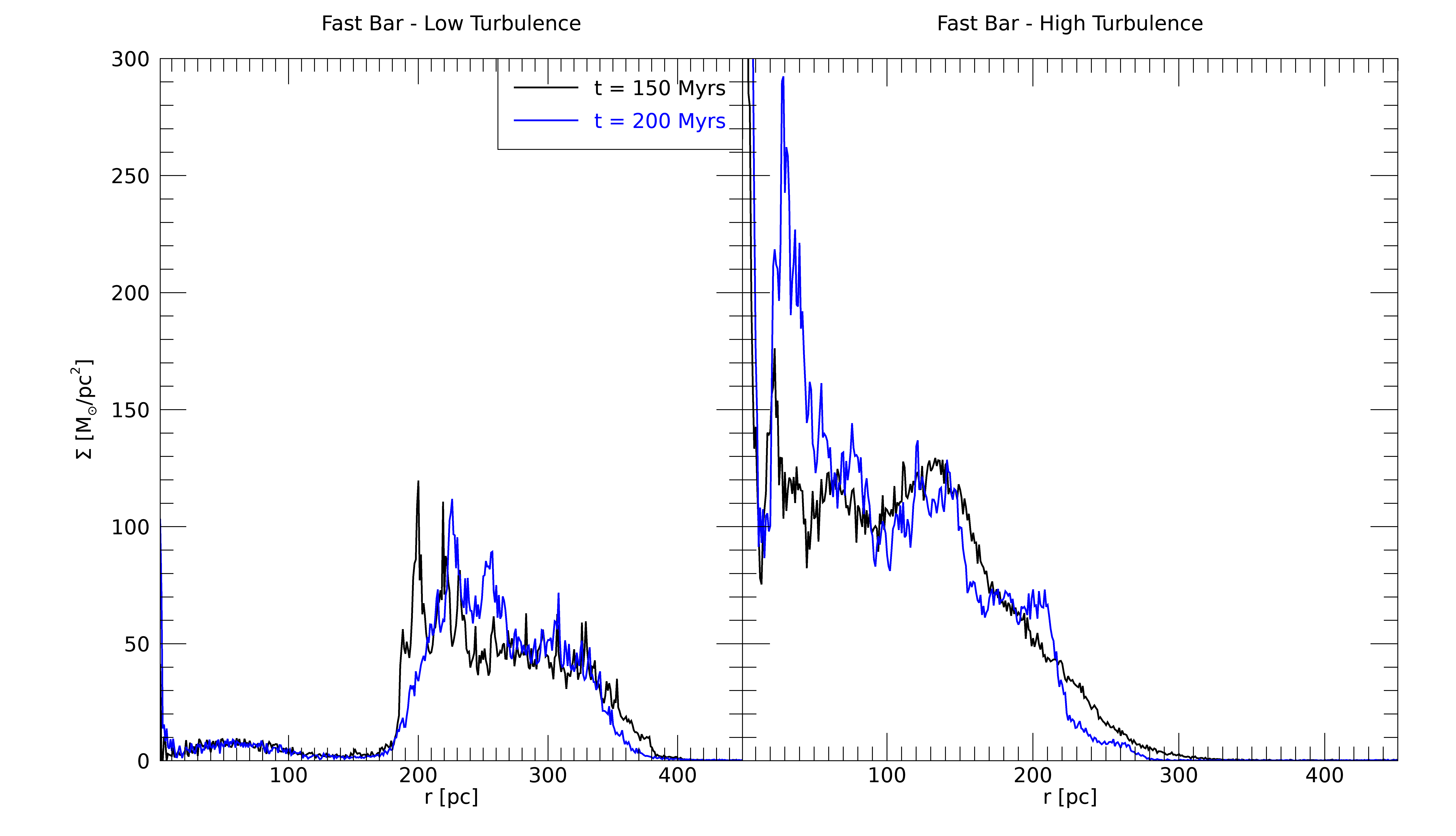}
\caption{Top: mass inside a radius of $r=400$~pc as a function of time. The black points represent the fast bar, no turbulence simulation (FBLSP). Similarly, the blue and red points represent the low turbulence (FBLT) and high turbulence (FBHT) simulations, respectively. We approximate the mass evolution after $150$~Myrs (i.e., after steady-state has been reached) as a straight line, and compute the slope, $\dot{M}$. Bottom: plots of surface density ($\Sigma$) versus radius for the FBLT (bottom left) and FBHT (bottom right) simulations. The black and blue lines correspond to $t=150$ and $200$~Myrs, respectively.  } 
\label{fig:fast_bar_plots}
\end{figure*}

To determine $\dot{M}$, we show in the top panel of Figure \ref{fig:slow_bar_plots} the mass inside a radius of $r=600$~pc (which is a radius large enough to contain the nuclear rings) of the slow-bar simulations with turbulence (low turbulence indicated in blue, high turbulence indicated in red). We consider only the time after $150$~Myrs, i.e., after the gas has settled into $X_1$ and $X_2$ orbits and reached a steady state, and fit the slopes as straight lines. We find an average mass inflow rate of $\dot{M}=0.05$ \msun/yr for the low turbulence run, and $\dot{M}=0.03$ \msun/yr for the high turbulence run. We also tabulate these results in Table \ref{table:visc_table}. For comparison, we also show in the top panel of Figure \ref{fig:slow_bar_plots} the corresponding values for the low sound speed, no turbulence run.

To calculate $\Sigma$, we plot in the bottom panel of Figure \ref{fig:slow_bar_plots} the surface density vs radius at 150 Myrs and 200 Myrs of the slow bar simulations with turbulence. We used the midpoint radius to calculate the average of both times, which is $r=500$~pc for the slow bar, low turbulence run (left panel of Figure \ref{fig:slow_bar_plots}), and $r=400$~pc for the slow bar, high turbulence run (right panel of Figure \ref{fig:slow_bar_plots}). The average values for the surface density at those radii for both runs are $\Sigma =60 $\msun/pc$^2$ and $\Sigma =80$ \msun/pc$^2$, respectively. Thus, using Equation \ref{eq:alpha2}, we compute $\nu_{turb}$ for both slow bar simulations with turbulence and present them in Table \ref{table:visc_table}, under the label ``Method 2''. Thus, we find that, for the slow bar (SB) simulations, the calculated values of $\nu_{turb}$ using Equations \ref{eq:alpha1} and \ref{eq:alpha2} (i.e., Methods 1 and 2) differ by less than a factor of two.

Similarly, Figure \ref{fig:fast_bar_plots} illustrates the same analysis as described for Figure \ref{fig:slow_bar_plots}, except for the fast bar simulations. However, to estimate $\dot{M}$, we consider in this case the mass inside a radius $r=400$~pc (enough to contain the smaller nuclear rings). Also, to calculate $\Sigma$, we use the average of both 150 and 200 Myrs at $r=300$~pc for the fast bar, low turbulence run (left panel of Figure \ref{fig:fast_bar_plots}) and $r=100$ pc for the fast bar, high turbulence run (right panel of Figure \ref{fig:fast_bar_plots}). We estimate the values of $\dot{M}$, $\Sigma$ and $\nu_{turb}$ and present them in Table \ref{table:visc_table}. We find that the calculated values of $\nu_{turb}$ using Equations \ref{eq:alpha1} and \ref{eq:alpha2} (i.e., Methods 1 and 2) to be different by about a factor of three in the case of the fast bar, high turbulence (FBHT) simulation. In the case of the fast bar, low turbulence (FBLT) simulation, the values of $\nu_{turb}$ differ by about a factor of 10. This can be explained by the fact that, as noted above, because the resulting nuclear ring in a fast bar simulation is relatively small, it receives a smaller fraction of the injected turbulent energy. This fact, combined with an already low turbulent injection rate (as is the case in the FBLT run), seems to not contribute significantly to the mass inflow compared to the fast bar simulation without turbulence (FBLSP, see the top panel of Figure \ref{fig:fast_bar_plots}).

To compare our calculated values of $\nu_{turb}$ to estimates of the viscosity derived from the characteristics of entire galaxies, we use the estimate from \cite{Lynden1974MNRAS.168..603L}:
\begin{equation}
    \nu = \frac{1}{3} c \frac{a}{f} \ ,
\end{equation}
where $c$ is the velocity dispersion, $a$ is a typical molecular cloud size, and $f$ is the filling fraction. Using the values $c=8$~km/s, $a=10$~pc and $f=0.1$, \cite{Lynden1974MNRAS.168..603L} estimated a turbulent viscosity of $\nu = 8\times10^{25}$~cm$^2$/s. The values of $\nu_{turb}$ we present in Table \ref{table:visc_table} are in good agreement with this estimation, especially for the slow bar simulations. 

Finally, the effects of viscosity on nuclear gaseous rings were explored also by \cite{Sormani2018aMNRAS.481....2S}, who found that nuclear rings spread over time, as expected.
We find qualitative similarities between their simulations and ours, especially for those with low turbulence (cf. figure 7 in \citealt{Sormani2018aMNRAS.481....2S}). However, we emphasize that unlike standard viscosity treatments, our turbulence method is capable of not only inducing an effective viscosity, but also producing the typical filamentary density structures that should be expected in the presence of turbulence.

\section{Summary and Conclusion}\label{sec:discussion}
As suggested by \citealt{Sormani2019MNRAS.488.4663S} (and references therein), the sizes of nuclear rings are generally controlled by: 1) the strength of the galactic bar, 2) the bar pattern speed, and 3) the sound speed. In this work, we explore the inclusion of turbulence, and compare it to the effects of thermal pressure due to low and high sound speeds (i.e., temperatures) in the absence of turbulence. We find that turbulence has the effect of spreading out nuclear rings, as well as promoting mass inflow to the center. Both of these effects are due to the viscosity generated by turbulence, as explored in \cite{Salas2019}.

The qualitative differences between the runs with and without turbulence are apparent in Figures \ref{fig:big_face_on} and \ref{fig:zoom_face_on}. Our turbulence treatment creates smaller and more dispersed rings than the low sound speed, no-turbulence tests. The creation of smaller rings was previously shown to depend on the speed of sound \citep[e.g.,][]{Kim2012ApJ...747...60K,Sormani2015aMNRAS.449.2421S,Sormani2018aMNRAS.481....2S}. In particular, the nuclear rings shrink in size with increasing sound speed and viscosity. Our turbulence method creates a turbulent viscosity, which has been shown to promote the transfer of angular momentum and enhance the rate of inward migration \citep{Wang2009ApJ...701L...7W,Hobbs2011MNRAS.413.2633H}, thus creating a smaller and more dispersed ring. 

This is especially true for the fast bar, high turbulence (FBHT) simulation, which has a ring that is completely filled in. This occurs because of the large turbulence injection rate in combination with a fast bar pattern speed. This is consistent with prior work showing that turbulence enhances inflow rates towards supermassive black holes (SMBHs). In particular, simulations by \cite{Tress2020MNRAS.499.4455T} showed that SN feedback drives gas from the CMZ into the Circumnuclear Disk located in the central few parsecs. This is because stellar feedback associated with episodes of star formation activity in the CMZ can stochastically launch parcels of gas towards the center (e.g., \citealt{Davies2007ApJ...671.1388D}). 
Other authors have suggested that SN feedback and supersonic turbulence inside accretion disks (e.g., \citealt{Wang2009ApJ...701L...7W}) can promote accretion onto SMBHs by enhancing angular momentum transfer (e.g., \citealt{Collin2008A&A...477..419C,Chen2009ApJ...695L.130C,Palou2020A&A...644A..72P}). This effect was confirmed also by \cite{Dinh2021}, which used the turbulence algorithm presented here to study the effects of turbulence on the evolution of the CND, and among their results, found that our turbulence method enhances accretion rates from the CND toward the central black hole.

Finally, it is worth noting that the ring produced by the FBHT test is similar, in terms of shape and mass, to that of \cite{Kim2011ApJ...735L..11K} (cf. their figure 1). This is not entirely surprising since our FBHT simulation uses the same gravitational potential, the same value for the bar pattern speed, and same initial mass of the large-scale disk as in their study. The main difference between the two simulations is the mechanism for turbulence driving. \cite{Kim2011ApJ...735L..11K} used a model of stellar feedback in which SN are simulated by injecting thermal energy into surrounding SPH particles. Our turbulence module injects kinetic energy in the form of a velocity field with a $k^{-4}$ power spectrum to all particles. This comparison shows that our turbulence driving method is comparable to standard methods of SN feedback in driving turbulence. However, turbulence in the Galaxy, and particularly in the CMZ, does not necessarily come only from SN feedback. In fact, the dominant source of turbulence in the CMZ has not been yet conclusively identified \citep{Kruijssen2014MNRAS.440.3370K}. Turbulence is produced by the interplay of many sources: SN blasts, gas instabilities, stellar winds, magnetic fields which produce hydromagnetic waves and MHD instabilities, etc., all which work in different scales and combine to create a turbulence spectrum that has been reported in the literature to approximately obey a power law \citep[e.g.,][]{Elmegreen2004ARA&A..42..211E}. Hence, the turbulence injection mechanism used here can account for many sources of turbulence within the adopted range of scales. 

In this work we focused only on a limited range of injection scales, as well as two limiting cases for the strength of the turbulence injection. Future simulations could focus on a more heuristic exploration of the turbulence parameters and turbulence scales, as well as a more refined gravitational potential of the inner Galaxy. Such simulations could be used to construct improved models that can be compared directly and statistically with the observed density and velocity dispersions in the CMZ. Furthermore, the turbulence method described here can be adapted to other SPH and grid codes, some of which could offer more elaborate techniques to capture the physics of the CMZ more accurately. Thus, the results of this paper indicate that our turbulence driving module is a promising way to model turbulence in this complex environment. 

\section*{Acknowledgments}
J.M.S would like to thank Sofia G. Gallego and Sungsoo S. Kim for their help in implementing the external gravitational potential, Diederik Kruijssen for helpful discussions about Galactic dynamics, and Blakesley Burkhart for insightful comments and discussions about hydrodynamics and turbulence. 
This material is based upon work supported by the National Science Foundation Graduate Research Fellowship Program under Grant No. DGE-1144087. This work used computational and storage services associated with the Hoffman2 Shared Cluster provided by UCLA Institute for Digital Research and Education's Research Technology Group. This work also used the Extreme Science and Engineering Discovery Environment (XSEDE) Comet at the San Diego Supercomputer Center at UC San Diego through allocation TG-AST180051. XSEDE is supported by National Science Foundation grant number ACI-1548562. SN acknowledges the partial support of NASA grant No. 80NSSC20K0500 and thanks Howard and Astrid Preston for their generous support.\\
{\bf Software:}
Figures \ref{fig:big_face_on} and \ref{fig:zoom_face_on} were done using the SPH visualization software \splash\ \citep{Price2007}.
We used Gadget2 \citep{Springel2005} to build our turbulence method. The version of the code that includes our turbulence routine can be found at \href{https://github.com/jesusms007/turbulence}{https://github.com/jesusms007/turbulence}.

\bibliographystyle{aasjournal}
\bibliography{paper1_ref}


\end{document}